\documentclass[aps,superscriptaddress,floats,onecolumn]{revtex4-1}
\usepackage{bm}

\usepackage{amssymb}
\usepackage{amsmath}
\usepackage{color}
\usepackage{natbib}
\newcommand\beq{\begin{equation}}
\newcommand\eeq{\end{equation}}
\newcommand\bea{\begin{eqnarray}}
\newcommand\eea{\end{eqnarray}}

\newcommand{\ra}{\rangle}
\newcommand{\la}{\langle}
\newcommand{\kan}{k[a_1,\dots,a_n]}
\newcommand{\np}{\mathbb{N}_0}

\newcommand{\N}{\mathbb{N}}
\newcommand{\Z}{\mathbb{Z}}
\newcommand{\ab}{{\bf a}}
\newcommand{\fs}{f_1,\dots,f_s}
\newcommand{\C}{\mathbb{C}}

\usepackage{graphicx}

\begin{document}

\title{Groebner basis methods for stationary solutions of a low-dimensional model for a shear flow}
\author{Marina Pausch}
\affiliation{Fachbereich Physik,
          Philipps-Universit\"at Marburg, D-35032 Marburg, Germany}
\author{Florian Grossmann}
\affiliation{Fachbereich Physik,
          Philipps-Universit\"at Marburg, D-35032 Marburg, Germany}
\author{Bruno Eckhardt}
\affiliation{Fachbereich Physik,
          Philipps-Universit\"at Marburg, D-35032 Marburg, Germany}
\author{Valery G. Romanovski}
\affiliation{CAMTP - Center for Applied Mathematics and Theoretical Physics\\
University of Maribor,
Krekova 2, Maribor SI-2000,  Slovenia}
\affiliation{
Faculty of Natural Science and Mathematics, University of Maribor\\
 Koro\v ska cesta 160,
   SI-2000 Maribor, Slovenia}
\date{\today}

\begin{abstract}
We use Groebner basis methods to extract all stationary solutions 
for the 9-mode shear flow model that is described in  
Moehlis et al, New J. Phys. {\bf 6} 54 (2004).  Using rational approximations
to irrational wave numbers and algebraic manipulation techniques we reduce
the problem of determining all stationary states to finding roots of a polynomial of order 30.
The coefficients differ by 30 powers of 10 so that algorithms for extended precision
are needed to extract the roots reliably. We find that there are 
eight stationary solutions consisting of two distinct states that each appear in four symmetry-related phases. We discuss extensions of these results for other flows.
\end{abstract}

\maketitle

\section{Introduction}
The transition to turbulence in several shear flows that do not show
a linear instability has been connected with the appearance of exact 
coherent states \cite{Faisst2003,Wedin2004,Eckhardt2007,Eckhardt2008b}. 
These states together with their heteroclinic connections 
form the scaffold on which the turbulent dynamics takes place 
\cite{Schmiegel1999,Eckhardt2008a,Gibson2008,Gibson2008a,Halcrow2009} and they 
dominate the statistical properties of the flow \cite{Schneider2007a,Eckhardt2008a,Kerswell2007}. 
In the case of plane Couette flow, a fluid confined between two parallel plates that move 
in opposite directions, the coherent states are of a particularly simple form because
of an up-down symmetry in the flow: they are stationary states \cite{Nagata1990,Clever1997,Wang2007,Schneider2008}. 
Because of the significance of these states it would be interesting to know how 
many fixed points there are. In reliable numerical simulations with their
thousands of degrees of freedom this amounts to finding zeros of the
equations of motion in very high dimensional spaces. 
The present approaches use suitably adapted Newton
methods \cite{Viswanath2007,Dijkstra2014} to find fixed points in the neighborhood of the initial conditions 
and therefore cover the state space only probabilistically: as the number of initial conditions tested
increases, the room for missed fixed points decreases, which makes their existence less 
likely but not impossible. Despite these sometimes very costly searches that require millions of initial conditions
one is not assured that no pocket with possibly new states is missed. Hence a different
approach to finding all states would be welcome.  We here discuss how
Groebner basis methods \cite{Weispfennig1993} can be applied to the problem at hand.

The original problem involves a partial differential equation, the Navier-Stokes equation, on 
a suitably defined domain with appropriate boundary conditions. For this infinite-dimensional
system only the simplest solutions can be obtained exactly (like the linear profile
for plane Couette flow) and all the ones relevant to the turbulence transition in shear 
flows are obtained numerically. However, due to the fact that the equation is dissipative
and that the fractal dimension of the turbulent state is finite (though increasing
with Reynolds number, see \cite{Ruelle1982,Temam1995}), 
projections onto finite dimensional subspaces give approximations that become more
accurate as the dimensions increase. Within such subspaces,
the equations are sufficiently simple and of a form that Groebner basis methods
can be used.

Specifically, an expansion of  the Navier-Stokes equation in appropriate divergence free
basis functions generically leads to coupled equations of the
form
\beq
\dot a_i = \sum_j d_{ij}a_j + \sum_{jk}\alpha_{i,jk}a_j a_k + f_i
\label{reduced}
\eeq
where the $a_i$ are the coefficients for the expansion of the velocity field. The precise
form of this representation (in Chandrasekhar modes or spectral-tau schemes or
others \cite{Orszag}) only affects the coupling coefficients in the equations and does not matter here.
The coefficients $\alpha$ cover the nonlinearity, the $d$'s are typically connected with
damping and the $f$'s are external forcings. Formally, the number of equations
is infinite because of the partial differential nature of the original equation, 
and good approximations converge to the exact result as the number of basis
functions increases. The conflicting requirement of a large number of modes for 
numerical accuracy and a small number of modes so that the Groebner basis
methods can be applied will be discussed in the concluding section.

The key feature of eq (\ref{reduced}) 
that enables the application of Groebner basis methods
is the quadratic coupling between amplitudes. 
As a consequence, in the generic case, that is, when the system (\ref{reduced}) 
does not have trajectories filled with singular points or, equivalently, the system of 
polynomials on the right hand side  of (\ref{reduced}) has only a finite number of zeros,    
there cannot be more than $2^N$ fixed points, where $N$ is the number of equations.
The actual number may be obtained by transforming the equations and extracting
a Groebner basis. Ultimately, this transforms the problem into one involving
the roots of a polynomial  in one of the variables, together with
equations that determine the remaining variables. The reduction in the number of
roots from the maximum of $2^N$ comes about because some of the roots may
be complex and hence meaningless, and because the final polynomial may
be of lower order. Specifically, for the case studied here, we can show
that there are only two sets of symmetry related roots.
\\ \indent
In the next section we focus the discussion on the 9-mode shear flow model derived
in \cite{Moehlis2004a,Moehlis2005}. Section 3 describes the Groebner basis search, and 
section 4 contains the results for  a specific aspect ratio. We conclude with a summary of the
results on the present model and an outlook of the implications for other
flows in section 5.

\section{The shear flow model}

The 9-mode model used here is obtained for a variation to the original plane Couette flow
problem that allows for a representation of the velocity fields in Fourier modes. To this end,
the no-slip boundary conditions at the top and bottom plate are replaced by free-slip
boundary conditions, and the driving is obtained from a sinusoidal body force. Furthermore,
the number of modes is severely truncated so that only nine modes survive. These modes have 
a physical interpretation, as explained in \cite{Moehlis2004a}:  two modes (amplitudes $a_1$ and
$a_9$) describe the 
laminar profile and its deformation due to other modes, one mode ($a_3$) describes downstream
vortices which are important for the extraction of energy from the laminar profile, and another mode
($a_2$) then captures the streaks that are induced by the vortices.  One pair of modes ($a_4$ and $a_5$) 
capture transverse motions that are needed in order to deform the flow in the downstream direction,
and another pair ($a_6$ and $a_7$) describes normal vortices. Finally, there is
a 3-d mode ($a_8$) that provides important couplings between the modes.  For
the functional form of the 3-d velocity fields we refer to \cite{Moehlis2004a}.
\\ \indent
The system has three parameters. The control parameter that we will use later on is the
Reynolds number $Re$. Two geometrical parameters $\alpha$ and $\gamma$ 
describe the length $\pi d/\alpha$ and the width $\pi d/\gamma$ of the domains 
(the height is fixed by $d=2$ with our choice of units).  
We study the system for the parameter values
of the larger of the two domains discussed in \cite{Moehlis2004a,Moehlis2005},
the so-called NBC domain with $\gamma=1$ and $\alpha=1/2$.

Then, the equations for the nine coefficients $a_1$, $\dots$, $a_9$ are given by:
\begin{eqnarray}
\frac{da_1}{dt} &=& \frac{\beta^2}{Re} -\frac{\beta^2}{Re} a_1 
- \sqrt{\frac{3}{2}} \left( \frac{\beta }{\kappa_{\alpha \beta \gamma}} a_6 a_8 
- \frac{\beta }{\kappa_{\beta \gamma}} a_2 a_3 \right),
\label{a1} 
\\
\frac{da_2}{dt} &=& -\left( \frac{4 \beta^2}{3} + 1 \right) 
\frac{a_2}{Re} 
+ \frac{5 \sqrt{2}}{3 \sqrt{3}} \frac{1}{\kappa_{\alpha \gamma}} a_4 a_6 - 
\frac{1}{\sqrt{6} \kappa_{\alpha \gamma}} a_5 a_7  
 - \frac{\beta}{2 \sqrt{6} \kappa_{\alpha \gamma} 
\kappa_{\alpha \beta \gamma}} a_5 a_8 \\* \nonumber
&& - \sqrt{\frac{3}{2}} \frac{\beta}{\kappa_{\beta \gamma}} (a_1 a_3 + a_3 a_9), 
\\
\frac{da_3}{dt} &=& -\frac{\beta^2 + 1}{Re} a_3 
+ \frac{1}{\sqrt{6}} \frac{ \beta}{\kappa_{\alpha \gamma} 
\kappa_{\beta \gamma}} (a_4 a_7 + a_5 a_6)   + \frac{7 \beta^2 -15}{4 \sqrt{6} \kappa_{\alpha \gamma} 
\kappa_{\beta \gamma} \kappa_{\alpha \beta \gamma}} a_4 a_8, 
\\
\frac{da_4}{dt} &=& -\frac{3 + 16 \beta^2}{12 Re} a_4 
- \frac{1}{2 \sqrt{6}} (a_1 a_5 +a_5 a_9)
- \frac{5}{6 \sqrt{6}} \frac{1}{\kappa_{\alpha \gamma}} a_2 a_6 
- \sqrt{\frac{3}{2}} \frac{\beta}{2 \kappa_{\alpha \gamma} 
\kappa_{\beta \gamma}} a_3 a_7\\* \nonumber
&& - \sqrt{\frac{3}{2}} 
\frac{\beta^2}{4 \kappa_{\alpha \gamma} 
\kappa_{\beta \gamma} \kappa_{\alpha \beta \gamma}} a_3 a_8 ,
\\
\frac{da_5}{dt} 
&=& -\frac{1 + 4 \beta^2}{4 Re} a_5 + \frac{1}{2 \sqrt{6}} (a_1 a_4 + a_4 a_9 )
+ \frac{1}{4 \sqrt{6} \kappa_{\alpha \gamma}} a_2 a_7 - \frac{\beta}{2 \sqrt{6} \kappa_{\alpha \gamma} 
\kappa_{\alpha \beta \gamma}} a_2 a_8 \\* \nonumber
&& + \frac{1}{\sqrt{6}} \frac{\beta}{\kappa_{\alpha \gamma} 
\kappa_{\beta \gamma}} a_3 a_6 
\\
\frac{da_6}{dt} &=& -\frac{15+ 16 \beta^2}{12 Re} a_6 
+ \frac{1}{2 \sqrt{6}} (a_1 a_7+ a_7 a_9)   + \sqrt{\frac{3}{2}} \frac{\beta}{\kappa_{\alpha \beta \gamma}} a_1 a_8  
- \frac{5}{2 \sqrt{6}} \frac{1}{\kappa_{\alpha \gamma}} a_2 a_4  \\* \nonumber
&&- \sqrt{\frac{2}{3}} \frac{\beta}{\kappa_{\alpha \gamma} 
\kappa_{\beta \gamma}} a_3 a_5 
+ \sqrt{\frac{3}{2}} \frac{\beta}{\kappa_{\alpha \beta \gamma}} a_8 a_9, 
\\
\frac{da_7}{dt} &=& -\frac{5 + 4 \beta^2}{4 Re} a_7 
- \frac{1}{2 \sqrt{6}} (a_1 a_6 + a_6 a_9) + \frac{3}{4 \sqrt{6}} \frac{1}{\kappa_{\alpha \gamma}} a_2 a_5 
+ \frac{1}{2 \sqrt{6}} \frac{\beta}{\kappa_{\alpha \gamma} 
\kappa_{\beta \gamma}} a_3 a_4, 
\\
\frac{da_8}{dt} &=& -\frac{5 + 4 \beta^2}{4 Re} a_8 
+ \frac{1}{\sqrt{6}} \frac{ \beta}{\kappa_{\alpha \gamma} 
\kappa_{\alpha \beta \gamma}} a_2 a_5  + \frac{15 - 4 \beta^2}{4 \sqrt{6} 
\kappa_{\alpha \gamma} \kappa_{\beta \gamma} \kappa_{\alpha \beta \gamma}} 
a_3 a_4,
\\
\frac{da_9}{dt} &=& -\frac{9 \beta^2}{Re} a_9 + \sqrt{\frac{3}{2}} 
\frac{\beta}{\kappa_{\beta \gamma}} a_2 a_3 
- \sqrt{\frac{3}{2}} \frac{\beta}{\kappa_{\alpha \beta \gamma}} 
a_6 a_8,
\label{a9}
\end{eqnarray}
where
\begin{eqnarray}
\beta &=& \frac{\pi}{2}, \\
\kappa_{\alpha \gamma} &=& \sqrt{\alpha^2 + \gamma^2} = \frac{\sqrt{5}}{2},\\
\kappa_{\beta \gamma} &=& \sqrt{\beta^2 + \gamma^2}= \frac{\sqrt{\pi^2 + 4}}{2},\\
\kappa_{\alpha \beta \gamma} &=& \sqrt{\alpha^2 + \beta^2 + \gamma^2}= \frac{\sqrt{\pi^2 + 5}}{2}.
\end{eqnarray}

The system (\ref{a1})-(\ref{a9}) possesses two discrete symmetries that are connected with 
the invariance of the flow under shifts along half the width and half the length, respectively, in the full domain, 
\cite{Moehlis2004a,Moehlis2005}:
\bea
T_{L_x/2} \vec a=(a_1, a_2, a_3, -a_4, -a_5, -a_6, -a_7, -a_8, a_9), \\ 
T_{L_z/2} \vec a=(a_1, -a_2, -a_3, a_4, a_5, -a_6, -a_7, -a_8, a_9),
\eea
They imply that for each solution $\vec a$ there exist another three 
symmetry-related solutions $T_{L_x/2} \vec a$,
$T_{L_z/2} \vec a$ and $ T_{L_x/2} T_{L_z/2} \vec a$.

The system of these nine modes and their couplings captures many features of shear flows 
and their dynamical behaviour \cite{Moehlis2004a,Moehlis2005}: 
the laminar profile is linearly stable for all Reynolds numbers,
the non-trivial state that appears first is not a periodic or in other ways simple state but
chaotic, and the chaotic dynamics are not persistent but transient. Moreover, the transition
scenario between simpler and more complicated states is similar to what has been seen
in pipe flow and other shear flows. 
Of course, it cannot capture the flow behaviour quantitatively
because of the small number of modes and hence poor resolution, and it differs from the 
full system in the boundary conditions.

\section{Groebner basis}
For fixed points $\dot a_i=0$, and the above equations reduce to 
a  polynomial   system of the form  
\begin{eqnarray}\label{s1} \nonumber
f_1(a_1,\dots,a_n)&=&0,\\
\vdots\qquad & &\\
f_m(a_1,\dots,a_n)&=&0,\nonumber
\end{eqnarray}
where $f_k$ are real 
polynomials and where the roots are the amplitudes of the stationary solutions. 
Even though such systems of  multivariate polynomials are pervasive in all areas 
of applied science and engineering,  no algorithmic  methods to solve systems like (\ref{s1})  
were known until the mid-sixties of the last 
century, when Bruno Buchberger \cite{Buchberger1965,Buchberger2006,Weispfennig1993} 
invented the theory of Groebner bases. They have since
become the cornerstone of modern computational algebra.  

We here summarize the notion of Groebner basis briefly and refer to 
\cite{Buchberger1965,Buchberger2006,Weispfennig1993,Arnold2003} for more details.
Let $k[a_1, \dots, a_n]$ denote the ring of polynomials of variables
  $a_1, \dots, a_n$ with coefficients in a  field $k$ and $I=\la f_1,f_2, \dots, f_s\ra$ denote  the ideal  generated by polynomials 
$f_1(a_1,\dots,a_n)$, $\ldots$,
$f_s(a_1,\dots,a_n)$, that is, the set of all sums
$
\{h_1 f_1+h_2 f_2+\dots + h_s f_s\},
$
where $f_k, h_k$ are polynomials.

A Groebner basis  of $I$ requires an ordering of the monomials of $\kan$, and different orderings give
different results. The two most commonly used term orders are the lexicographic order  and the 
degree reverse lexicographic order, defined as follows.
Let $\mu = (\mu_1, \dots,\ \mu_n)$ and
$\nu = (\nu_1, \dots, \nu_n)$ be elements of $\np^n$ ($\np=\N\cup 0$).
We say that 
     $\mu \succ_{\rm lex} \nu$  with respect to the lexicographic order if and only if, reading left to right,
     the first nonzero entry in the $n$-tuple
     $\mu - \nu \in \Z^n$ is positive.
Similarly, we say that 
     $\mu \succ_{\rm degrev} \nu$  with respect to the  degree reverse  lexicographic order 
if and only if 
     $
     |\mu| = \sum_{j=1}^n \mu_j> |\nu| = \sum_{j=1}^n \nu_j
     $
       or
     $
     |\mu| = |\nu| \ 
     $ 
and, reading right to left, the first nonzero entry in the   $n$-tuple
     $\mu - \nu \in \Z^n$ is negative.

Furthermore, we introduce the abbreviated notation $\ab^{\sigma}$  for 
 the monomial $a_1^{\sigma_1}a_2^{\sigma_2}\cdots  a_n^{\sigma_n}$ with $\sigma\in \np$.
 Fixing a term order  on $\kan$, any  $f \in \kan$ may then be written
in the \emph{standard form} with respect to the order,
\begin{equation}\label{standard}
f = \alpha_1 \ab^{\mu_1} + \alpha_2 \ab^{\mu_2} + \dots + \alpha_s \ab^{\mu_s},
\end{equation}
where  $\mu_i \ne \mu_j$ for $i \ne j$
and $1 \le i,j \le s$, and where, with respect to the specified term order,
$\mu_1 \succ \mu_2 \succ \cdots \succ \mu_s$.
The \emph{leading term}\index{term!leading} $LT(f)$ of $f$ is the term
     $LT(f) = \alpha_1 \ab^{\mu_1}$.

Let $f$ and $g$ be from  $\kan$ with $LT(f) = \alpha \ab^\mu$ and
$LT(g) = \beta \ab^\nu$.
The \emph{least common multiple}
of $\ab^\mu$ and $\ab^\nu$, denoted
$LCM(\ab^\mu,\ab^\nu)$, is the monomial
$\ab^\sigma = a_1^{\sigma_1} \cdots a_n^{\sigma_n}$ such that
$\sigma_j = \max(\mu_j, \nu_j)$, $1 \le j \le n$, and  the
\emph{$S$-polynomial} of $f$ and $g$ is the polynomial
\[
S(f,g)=\frac{\ab^\sigma}{LT(f)}f -\frac{\ab^\sigma}{LT(g)} g.
\]

The following algorithm due to Buchberger \cite{Buchberger1965}  produces a Groebner basis $G=\{g_1,\dots,
g_s\}$ for the ideal  $I=\la\fs \ra \in
\kan $:
%\medskip
\begin{itemize}
%\noindent 
\item[]Step 1: $G := \{ \fs \}$.\\
\item[]Step 2:  For each pair $g_i, g_j \in G$, $i \ne j$, compute the
       $S$-polynomial $S(g_i, g_j)$ and  compute the remainder $r_{ij}$
of the division  $S(g_i, g_j)$ by $ G$.  \\
\item[]Step 3: If
all $r_{ij}$ are equal to zero, output $G$, \\
else \\
add all nonzero $r_{ij}$ to $G$ and return to Step 2.
\end{itemize}

It is clear that the set of solutions of the system 
$$g_1=\dots=g_s=0$$
is the same as the set of solutions of the original system (\ref{s1}).

The convergence criterion in the last step requires that the coefficients vanish
exactly. This can only be met if the coefficients of the polynomials are integers 
or rational numbers (which become integers once the equation is multiplied 
with the least common multiple of all denominators). 

A Groebner basis  $G = \{ g_1, \dots, g_m \}$ is called \emph{reduced} 
 if for all $i$, $1 \le i \le m$,
the coefficient of the leading term is 1 and no term of $g_i$ is divisible 
by any  $LT(g_j)$ for $j \ne i$.
It is well known (see e.g. \cite{Cox1997,Romanovski2009})  that system (\ref{s1})
has a solution over the field of complex numbers  if and only if
the reduced Groebner basis $G$ for $\la \fs \ra$ with respect to any term order on
$\C[a_1, \dots, a_n]$ is different from $\{ 1 \}$.
The Groebner basis  theory allows  to find all solutions of system 
(\ref{s1}) in the case the system has only a finite number of solutions. 
In such case 
a Groebner basis with respect to 
the lexicographic order  is always in a ``triangular" form.

If system (\ref{s1}) has only a finite number of solutions, then any reduced Groebner basis  with 
respect to a lexicographic  order  \emph{must} contain a polynomial in one variable, say, 
$g_1(a_1)$. Then, there is a group of polynomials in the Groebner basis  depending on this variable 
and one more variable, say, $g_2(a_1,a_2), \dots, g_t(a_1,a_2) $ etc. Thus, we  first solve 
(perhaps numerically) the equation $g_1(a_1)=0$. Then, for each solution $a_1^*$ of   $g_1(a_1)=0$
we find the solutions of  $g_2(a_1^*,a_2)= \dots= g_t(a_1^*,a_2)=0, $ 
which is a system of polynomials in a single variable $a_2$. Continuing the process we obtain 
in this way all solutions of our system (\ref{s1}). Therefore, in the   case   of a finite number of 
solutions, Groebner basis computations provide the  complete solution to the problem
(see e.g. \cite{Cox1997} for more details).  However, this theoretical result is in practice clouded by 
considerable technical difficulties, since calculations of Groebner bases, especially with respect to  
lexicographic orders, require tremendous computational resources, as the 
size of the coefficients of $S$-polynomials  grows exponentially during the execution of the algorithms.

Therefore, the computation of Groebner bases requires powerful computers and efficient 
program packages.  Nowadays all major computer algebra systems 
(Mathematica, Maple, REDUCE, SINGULAR, Macaulay  and  many others) have 
routines to compute Groebner bases, which use much more efficient algorithms than
the original ones  described above.  To our experience one of the most efficient software tools for 
this purpose is the computer algebra system { SINGULAR} \cite{Greuel2005}  because of its rich
functionality and  high performance in implementation of constructive algorithms. 
In our study of the system  (\ref{a1})-(\ref{a9}) we were able to complete the computation  over the field of 
the rational numbers. 

As mentioned, Groebner bases cannot be calculated over real numbers, so all irrational numbers
have to be approximated by rational ones. The Reynolds numbers we study are between 200 and 1000
and are taken to be integers. The geometrical factors and the coupling contain irrational numbers that
have to be approximated by ratios. We noticed that poor approximations gave results incompatible
with numerical simulations of the full equations, so we finally settled for the representations
$$
\alpha=1/2, \quad \beta=157/100 \approx \frac{\pi}{2}, \quad \gamma=1
$$ 
and 
$$
\sqrt{3/2} \approx 564597/460992, \quad \sqrt{2/3} \approx 837390/1025589,  
$$ 
$$
\sqrt{6} \approx 1590187/649191,
$$
$$
\kappa_{\alpha\gamma} \approx 1934705/1730453, \quad \kappa_{\beta\gamma} \approx 1441913/774629, 
$$
$$
\kappa_{\alpha\beta\gamma} \approx 334287/173438.
$$

\section{Results for the shear flow model}
The Groebner bases of the 9-mode shear flow model were  determined for Reynolds numbers 
200, 300, 350, 400, 500 and  1000 (each Reynolds number required an independent determination of the
basis). 
In each case  we found a basis consisting of 11 polynomials, with one of them,  $j_1(a_9)$, depending on $a_9$ only,
and the others containing $a_9$ and one or more of the other coefficients.
Therefore, the roots of $j_1$ define the possible values of $a_9$ and the other equations give
the values for the other coefficients, or, if no real solutions can be found, eliminate a particular 
root of $j_1$. 
The fixed points of the 9-mode shear flow model are
mapped equivalently to the roots of polynomials on the right hand side of 
(\ref{reduced}) and the roots are determined by computing the reduced Groebner basis of 
the polynomials using the lexicographic ordering 
with $a_1\succ a_2\succ a_3\succ a_4\succ  a_5\succ a_6\succ a_7\succ a_8\succ a_9$.
The Groebner basis found for the 9-mode shear flow model generally has the following structure:
\begin{align}
%\beq
j_1(a_9)\;&=\;a_9\; p_1(a_9)\\[0.2cm]
%\eeq
%\beq
j_2(a_8,a_9)\;&=\;a_8\; p_2(a_9)\\[0.2cm]
%\eeq
%\beq
j_3(a_8,a_9)\;&=\;c_3\; (a_8)^2\; +\; p_3(a_9)\\[0.2cm]
%\eeq
%\beq
j_4(a_7,a_8,a_9)\;&=\;c_4 \;a_7\; +\; a_8\; +\; p_4(a_9)\\[0.2cm]
%\eeq
%\beq
j_5(a_6,a_8,a_9)\;&=\;c_5\; a_6\; +\; a_8\; p_5(a_9)\\[0.2cm]
%\eeq
%\beq
j_6(a_5,a_9)\;&=\; a_5 \;p_6(a_9)\\[0.2cm]
%\eeq
%\beq
j_7(a_5,a_9)\;&=c_7\; (a_5)^2\; +\; p_7(a_9)\\[0.2cm]
%\eeq
%\beq
j_8(a_4,a_5,a_9)\;&=\;c_8\; a_4\; + \;a_5 \;a_8\; p_8(a_9)\\[0.2cm]
%\eeq
%\beq
j_9(a_3,a_5,a_8,a_9)\; &=\; c_9\; a_3\; +\; a_5\; a_8\; p_9(a_9)\\[0.2cm]
%\eeq
%\beq
j_{10}(a_2,a_5,a_8,a_9)\;&=\;c_{10}\; a_2\; + \;a_5\; a_8\; p_{10}(a_9)\\[0.2cm]
%\eeq
j_{11}(a_1,a_9)\;&= \;a_1\;-\;9\; a_9\;-\;1 
\end{align}
with coefficients $c_3$ to $c_{10}$ and the polynomials 
\beq
p_\nu(a_9)= \sum_{\mu=0}^{30} \alpha_{\nu\mu} (a_9)^\mu.
\eeq
The coefficients $c_\nu$ and the polynomial coefficients $\alpha_{\nu\mu}$ depend on the 
Reynolds number, but the structure of the Groebner basis does not.

These coefficients are integer numbers of enormous size and varying signs.
For instance, for $Re=400$ the coefficients $\alpha_{2j}$ have more than 1300 digits and 
vary over up to 29 orders of magnitude,
\beq
\left| \frac{\alpha_{2,25}}{\alpha_{2,0}} \right| \approx \frac{1.23\cdot 10^{1315} }{4.98 \cdot 10^{1285}} \approx 2.5 \cdot 10^{29}.
\eeq
Together with alternating signs this causes enormous cancellations and considerable numerical difficulties.
This applies in particular in the regions of most interest, where the roots of the 
polynomials are located, e.g. $p_2(a_9)$ with  $a_9\in [-1,1]$. For example, $a_9=10^{-1}$
results in 
$(a_9)^{30}=10^{-30}$, 
which is a variation by $30$ orders of magnitude in the powers of the variable alone.
In order to overcome this problem arbitrary precision algorithms of  the "Class library for numbers" (CLN) 
are used to compute the functional values of the polynomials using the Horner algorithm.
The number of reliable digits of CLN floating point numbers can be tuned to any desired 
accuracy \cite{cln_manual}. 
The implementation of elementary arithmetics of the CLN is remarkably fast, so that 
root finding for the polynomial $p_2(a_9)$ by Newton's method is just a matter of seconds even when 
numbers with thousands of digits have to be handled. 

For our case, where the coefficients vary by about 30 orders of magnitude, we found that 
55 significant digits are sufficient to determine the real roots of the 9-mode model up 
to a relative error of approximately $10^{-4}$ using Newtons method. This is of the same  order 
as the error introduced by representing the real coefficients of the 9-mode model by rationals
needed for the Groebner basis.
To be on the safe side for all parameter values, we worked with 100 digits, 
so that the roots could be determined with at least 30 reliable digits.

\begin{figure}
\includegraphics[width=6cm,angle=270]{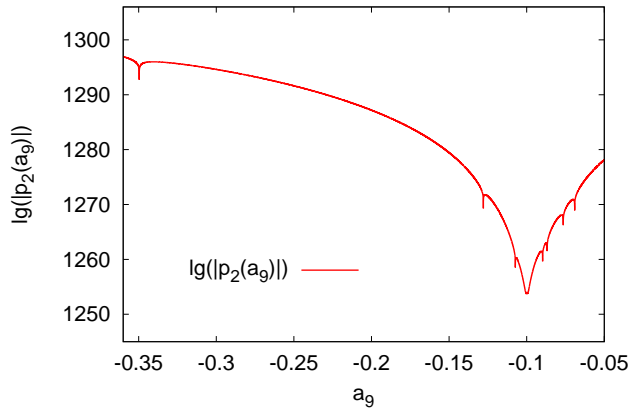}
\caption[]{The decadic logarithm of the value of the polynomial $p_2(a_9)$ reveals roots as
logarithmic singularities, which are easily identified in the plot. From left to right there are 3 roots followed
by an extremum and another four roots.}
\label{fig_p2_survey}
\end{figure}

It is not only the range that is large, but the absolute numbers as well.
The coefficients in the Groebner basis can be as large as $10^{27000}$.
For instance, when we plot the polynominal $p_2(a_9)$ in  Figure \ref{fig_p2_survey},
we use the decadic logarithm $\lg|p_2(a_9)|$ to cover values up to $\mathcal{O}(10^{1300})$.
The roots of the polynomial result in logarithmic singularities in this 
diagram, which are easily identified in the otherwise smooth function.

The polynomial $p_2(a_9)$ and thus also the polynomial $j_2(a_8, a_9)$ 
possess exactly 8 roots in variable $a_9$ indicated
by $a'_{9,s}$ with $s\in\{1,2,\ldots,8\}$. The polynomials $p_1$ and $p_6$ coincide with $p_2$.
From polynomials $j_3(a_8,{a'_{9,s}})$ and $j_7(a_5,{a'_{9,s}})$ necessarily follow values for the roots 
$a'_{8,s,t}$ and $a'_{5,s,t}$ of these polynomials with $t\in\{1,2\}$, the multiplicity is due to the free choice in sign.
The quadratic occurrence of $a_8$ and $a_5$ in these polynomials suggests that there are up to four times eight 
real roots
\beq
a'_{8,s,t}=\pm \sqrt{-\frac{p_3(a'_{9,s})}{c_3}}
\label{eqn_a8_explicit}
\eeq
and
\beq
a'_{5,s,t}=\pm \sqrt{-\frac{p_7(a'_{9,s})}{c_7}}
\label{eqn_a5_explicit}
\eeq
in these variables.
Interestingly, not all roots $a'_{9,s}$ result in real values of $a'_{8,s,t}$ and $a'_{5,s,t}$, which indicates that 
not all roots of $j_2(a_8, a_9)$ belong to a set of common roots $\{a_{\mu,r}\}$ of the Groebner basis $j_\nu(\{a_{\mu}\})$.
Table \ref{tab_groebner_root_hierachy} shows the hierarchy of resulting roots of $j_2$, $j_3$ and $j_7$ followed by their
implications on the Groebner basis.

\begin{table}
\caption{The roots $a'_{9,s}$ of polynomial $j_2(a_8, a_9)$ and the resulting values of $a'_{8,s,t}$ and $a'_{5,s,t}$
indicate that only two roots of $j_2(a_8, a_9)$ belong to a set of common roots of the Groebner basis.}
\begin{tabular}{ c | c | c | c | c}
s	&	$a'_{9,s}$	&	$a'_{8,s,t}$		&	$a'_{5,s,t}$		&	Root of Groebner basis\\
\hline
1	&	$-0.350$	&	$\pm 0.827$		&	$\not \in \mathbb{R}$	&	No\\
2	&	$-0.128$	&	$\pm 2.13$		&	$\not \in \mathbb{R}$	&	No\\
3	&	$-0.107$	&	$\not \in \mathbb{R}$	&				&	No\\
4	&	$-0.0897$	&	$\pm 0.0631$		&	$\pm 0.0356$		&	Yes\\
5	&	$-0.0868$	&	$\not \in \mathbb{R}$	&				&	No\\
6	&	$-0.0764$	&	$\pm 0.0734$		&	$\pm 0.0771$		&	Yes\\
7	&	$-0.0691$	&	$\pm 2.25$		&	$\not \in \mathbb{R}$	&	No\\
8	&	$0.00924$	&	$\pm 0.00414$		&	$\not \in \mathbb{R}$	&	No\\
\\
\end{tabular}
\label{tab_groebner_root_hierachy}
\end{table}

The structure of the Groebner basis clearly reveals that all the remaining polynomials do not impose further restrictions 
on the two roots $a_{9,4}$ and $a_{9,6}$ such that each of these roots of $j_2(a_8, a_9)$ gives rise to a 
real root of the Groebner basis.
The remaining $j_\nu({a_\mu})$ can successively be solved for the remaining undetermined $a_\mu$, 
which are then uniquely determined once the signs of $a_8$ and $a_5$ have been chosen.

Table \ref{tab_groebner_basis_8_solutions} lists all eight roots of the Groebner basis of the 9-mode shear flow 
model at Reynolds number $Re=400$. The different roots of the Groebner basis follow from the signs of the 
roots $a'_{9,4}$ and $a'_{9,6}$ of polynomial $j_2(a_8, a_9)$.
From the theory of Groebner bases discussed above these sets of variables are equivalent to fixed points of 
the 9-mode shear flow model, which will be investigated in the following.

\begin{table}
\caption[]{A list of all eight roots of the Groebner basis at $Re=400$ following from 
the roots $a'_{9,4}$ (top) and $a'_{9,6}$ (bottom)
of polynomial $j_2(a_8, a_9)$. A root of the Groebner basis consists of a column of signs multiplied 
to the column of values.
The four sets of signs follow from Equations (\ref{eqn_a8_explicit}) and (\ref{eqn_a5_explicit}).
The values of the $\{a_\mu\}$ have been computed much more accurately than displayed and have 
been truncated for this list.}
\begin{tabular}{ c | c | c | c | c | r}
\hline
\hline
$r$	&	1	&	2	&	3	&	4	&\\
\hline
$a'_{8,4}$ &	$>0$	&	$<0$	&	$>0$	&	$<0$	&\\
$a'_{5,4}$ &	$>0$	&	$>0$	&	$<0$	&	$<0$	&\\
\hline
$a_{9,r}$	&		&		&		&		& -0.0897\quad\!44635\quad\!37647\\
$a_{8,r}$	& 	+	&	-	&	+	&	-	& 0.0631\quad\!22811\quad\!60210\\
$a_{7,r}$	&	-	&	+	&	-	&	+	& 0.1794\quad\!15620\quad\!62289\\
$a_{6,r}$	&	+	&	-	&	+	&	-	& 0.1063\quad\!13203\quad\!79275\\
$a_{5,r}$	&	+	&	+	&	-	&	-	& 0.0356\quad\!35231\quad\!54453\\
$a_{4,r}$	& 	+	&	+	&	-	&	-	& 0.0168\quad\!03142\quad\!01677\\
$a_{3,r}$	& 	+	&	-	&	-	&	+	& 0.0345\quad\!33570\quad\!69306\\
$a_{2,r}$	& 	+	&	-	& 	-	&	+	& 0.0481\quad\!49018\quad\!20195\\
$a_{1,r}$	&	+	&	+	&	+	&	+	& 0.1922\quad\!98281\quad\!61171\\
\hline
\hline
r	&	5	&	6	&	7	&	8	&\\
\hline
$a'_{8,6}$ &	$>0$	&	$<0$	&	$>0$	&	$<0$	&\\
$a'_{5,6}$ &	$>0$	&	$>0$	&	$<0$	&	$<0$	&\\
\hline
$a_{9,r}$	&		&		&		&		& -0.0764\quad\!45389\quad\!20741\\
$a_{8,r}$	& 	+	&	-	&	+	&	-	& 0.0734\quad\!20744\quad\!07881\\
$a_{7,r}$	&	-	&	+	&	-	&	+	& 0.2989\quad\!49154\quad\!13129\\
$a_{6,r}$	&	+	&	-	&	+	&	-	& 0.0711\quad\!93885\quad\!61847\\
$a_{5,r}$	&	+	&	+	&	-	&	-	& 0.0771\quad\!37476\quad\!48830\\
$a_{4,r}$	& 	+	&	+	&	-	&	-	& 0.0157\quad\!42387\quad\!67594\\
$a_{3,r}$	& 	+	&	-	&	-	&	+	& 0.0355\quad\!82985\quad\!62015\\
$a_{2,r}$	& 	+	&	-	& 	-	&	+	& 0.0265\quad\!26551\quad\!11492\\
$a_{1,r}$	&	+	&	+	&	+	&	+	& 0.3119\quad\!91497\quad\!13324\\
\hline
\end{tabular}
\label{tab_groebner_basis_8_solutions}
\end{table}

Given the numerical difficulties with the polynomials we verified the roots by substituting them into the
equations of motion (with ordinary double precision floating point arithmetic). We find that all roots of the
Groebner basis listed in Table \ref{tab_groebner_basis_8_solutions} are indeed fixed points of
the 9-mode shear flow model. 
The different sign combinations mirror the symmetry relations of the amplitude equations.

\begin{figure}
\includegraphics[width=8cm,angle=0]{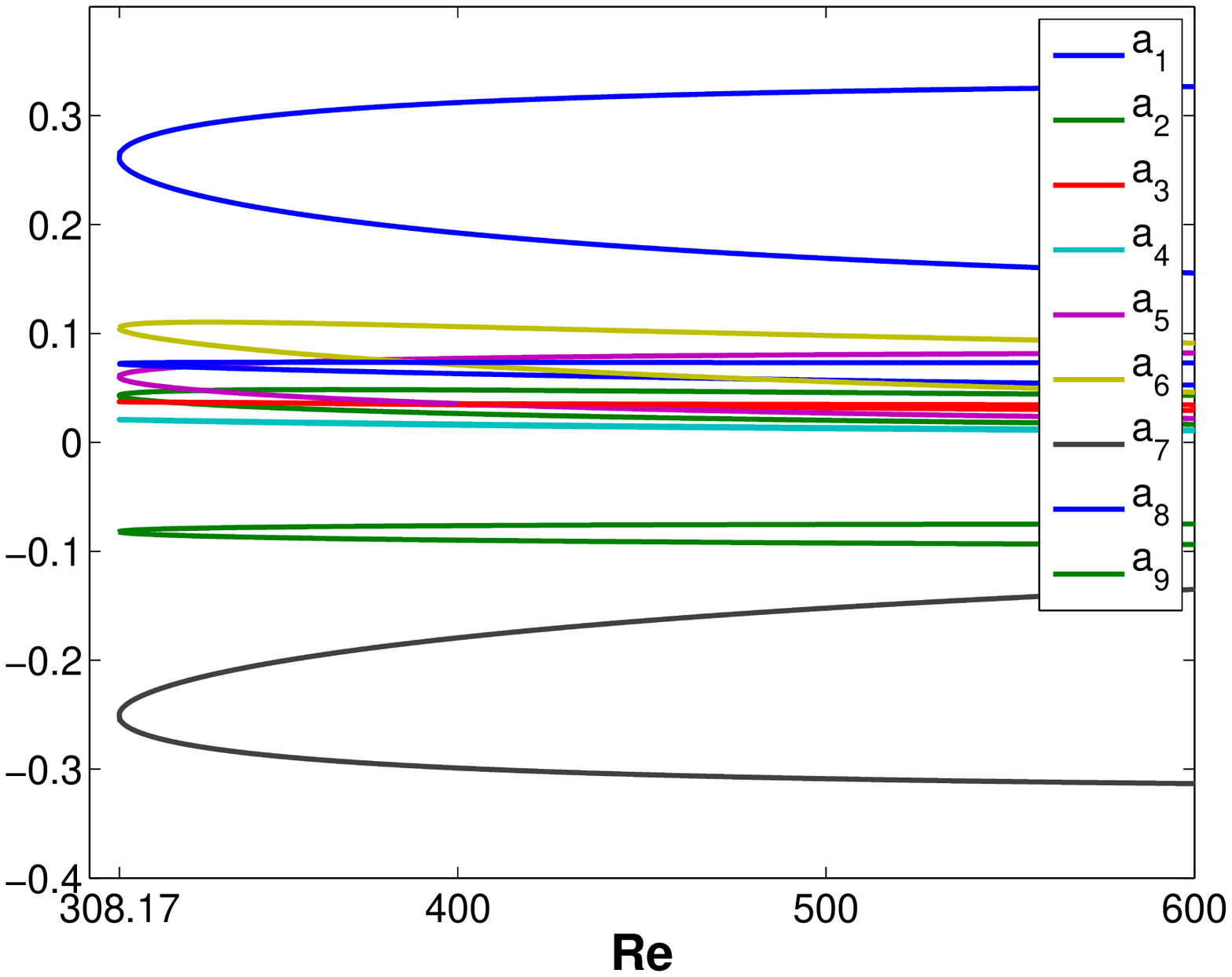}
\caption[]{The bifurcation diagram shows the merging of the fixed points $\{a_{\mu,1}\}$ and $\{a_{\mu,5}\}$ of the 9-mode shear flow model
at the critical Reynolds number $Re_{crit}=308.17$.}
\label{fig_bifurcation_diagram}
\end{figure}

The physical states corresponding to the fixed points of the 9-mode shear flow model are functions of the
Reynolds number.  It is relatively straightforward to track the roots, once they have been determined for
one Reynolds number, to different Re. A gradual decrease of the Reynolds number shows that the fixed 
points do not vary much with the Reynolds number until it reaches $Re\approx 308.17$, where the coefficients
quickly approach each other and annihilate in an inverse saddle node bifurcation, 
as shown in Figure \ref{fig_bifurcation_diagram}. 

As the Groebner basis has been determined for $Re=400$ it is guaranteed that there exist just the eight fixed 
points found for exactly this Reynolds number. In principle there could exist more fixed points for other Re, not connected
to the ones identified through bifurcations. We therefore computed Groebner bases 
for $Re = \{200,300,350,500,1000\}$. For the lowest two values $Re=200$ and $Re=300$ the Groebner 
bases do not posses any roots besides the trivial one, so there are no fixed points and stationary solutions in this flow regime. 
For all other Reynolds numbers there exist exactly eight roots of the Groebner basis as 
is the case for $Re=400$ and the corresponding fixed points are identical to those obtained from root tracking. 
From these checks we conclude that there exist only eight fixed points of the 9-mode shear flow model in 
this range of Reynolds numbers.

To conclude this section, we display the velocity field of the fixed point that has been obtained by these
methods, at the point of bifurcation at $Re=308.17$. One clearly notes the pair of vortices and the modulations 
in the downstream velocity, the streaks. As one moves above the critical Reynolds number for the
bifurcation, the upper and lower branch differ slightly in the velocity pattern from the one at the point
of bifurcation, but the vortices and streaks remain prominent features.

\begin{figure}
\includegraphics[width=8cm,angle=0]{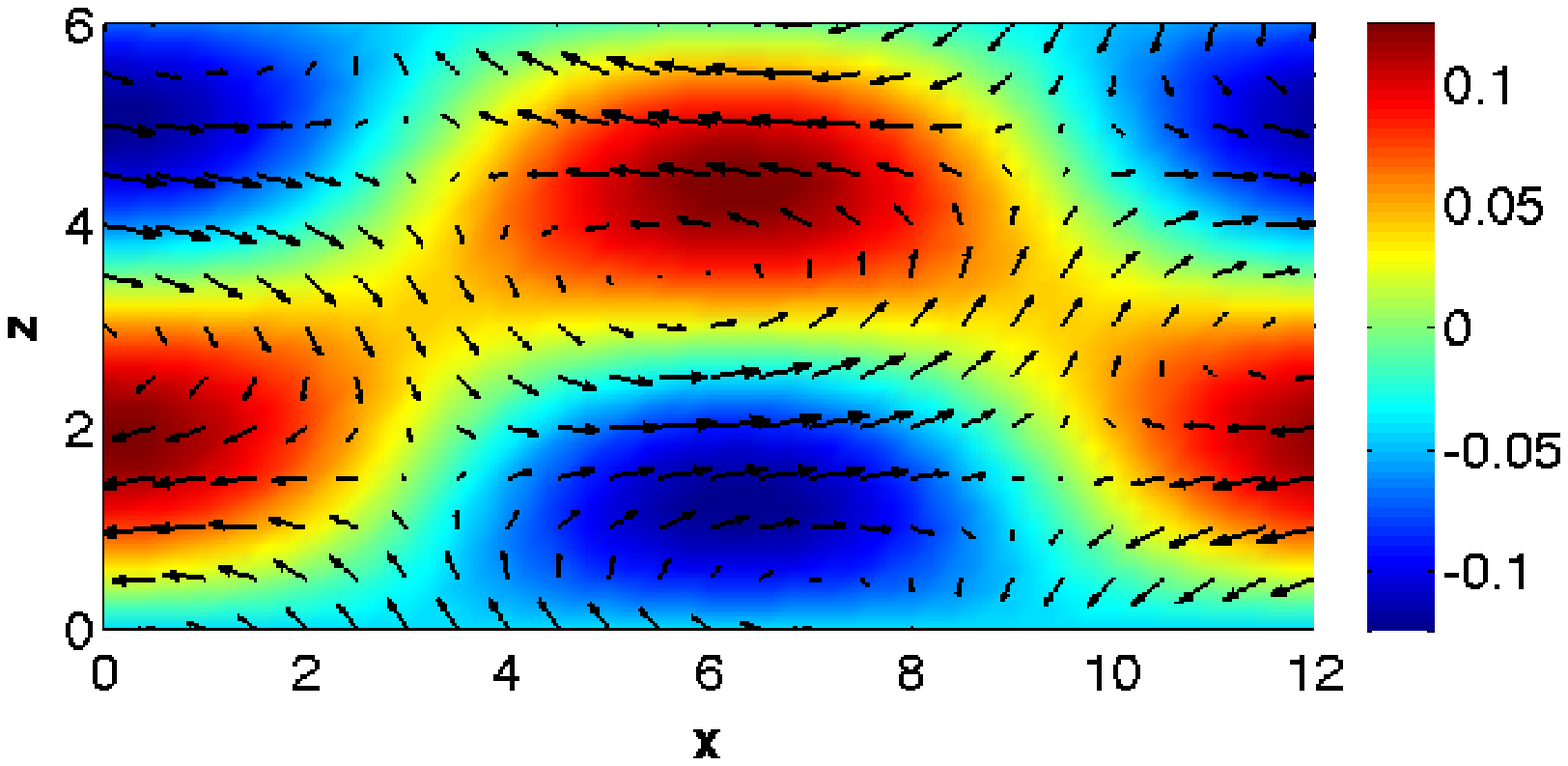}
\includegraphics[width=8cm,angle=0]{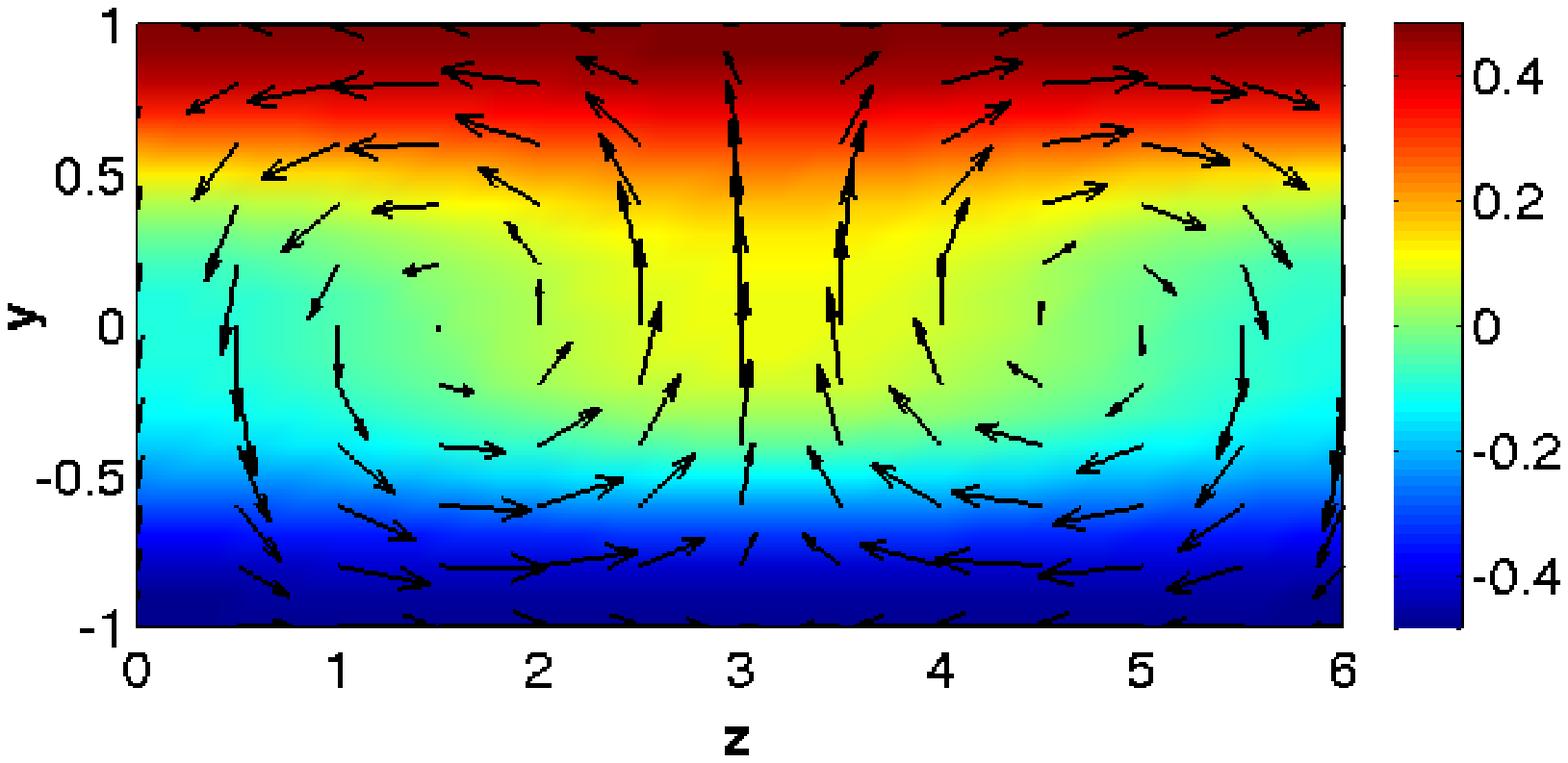}
\caption{Velocity field of one fixed point at the saddle-node bifurcation at $Re\approx 308.17$. Left image: Flow in the midplane of the domain ($y=0$). The color indicates the velocity in $y$-direction and the arrows indicate the flow in the $(x,z)$-plane. Right image: The color and the arrows indicate the averaged flow in $x$-direction and the flow in the $(y,z)$-plane, respectively.}
\label{fig_flowfield_bif}
\end{figure}

\section{Conclusions}
The Groebner basis analysis has confirmed that the two sets of four symmetry related fixed points found
numerically in \cite{Moehlis2005} are indeed the only ones in this setting. They appear
in a saddle node bifurcation near $Re=308.17$ and continue to exist for all larger
Reynolds numbers. The absence of further fixed points at higher Reynolds numbers
is somewhat unexpected, especially in view of the high degree of the polynominal and the
large number of additional states found in direct numerical simulations of the
full system \cite{Schmiegel1999,Gibson2008a}. However, all these states have 
a more complicated spatial structure that cannot be captured by the simple
Fourier modes included in the present model. This suggests that more states
can only be found if the model is extended to include additional spatial degrees 
of freedom that can become dynamically active.

As in other cases, the dynamics of the Groebner basis calculation remains mysterious
and unpredictable. On top of that, the calculation of roots of polynomials of
order 30 with coefficients O(1) near the origin turns out to be unexpectedly
troubling, requiring very high precision arithmetic. The dependence on the
parameters, in particular the aspect ratio and the Reynolds number seems to
be sufficiently well behaved, at least for the aspect ratios studied here. This also
suggests that the rational approximation of real quantities that is required for 
the Groebner analysis can be tolerated. 

Given the complexity of the calculation one would like to take advantage of as much
prior information as possible, and include, for instance, the discrete symmetries. 
The four-fold discrete degeneracy of the solutions shows up in the 
quadratic equations for $a_8$ and $a_5$ in equations (\ref{eqn_a8_explicit})
and (\ref{eqn_a5_explicit}), respectively. However, we are not aware of methods
that would allow to include this in a Groebner basis algorithm.

The existence of stationary solutions is a special feature of plane Couette flow,
connected with the discrete up-down symmetry of the velocity field 
\cite{Nagata1990,Clever1997,Wang2007}. Once this symmetry is broken the
states move downstream in the form of traveling waves. 
In more general cases, like plane Poiseuille flow or pipe flow, no stationary solutions can exist
and travelling waves are the simplest states that can appear \cite{Faisst2003,Wedin2004}. 
Since travelling waves become fixed points in a co-moving frame of reference, 
a similar analysis should be possible at the expense of yet another parameter, 
the phase speed of the solutions. 
Now the issue would be to find the number of solutions for a given Reynolds number
and all possible phase speeds. This requires either multiple transformations
for a set of phase speeds (similar to the scan of Reynolds numbers discussed
in section 2) or the analysis of a higher order system with the phase speed
as an additional parameter.

{\em Acknowledgements.} 
This work was  partly supported by the Deutsche Forschungsgemeinschaft
with Forschergruppe 1182. VR acknowledges  the support
by the Slovenian Research Agency  and  by the Transnational Access Programme at
RISC-Linz of the European Commission Framework 6 Programme for
Integrated Infrastructures Initiatives under the project SCIEnce (contract no. 026133).

\bibliographystyle{apsrev4-1}
\bibliography{library_groebner}

\end{document}